\documentclass[10pt]{iopart}
\usepackage{graphicx} 
\usepackage{xcolor}
\usepackage{upgreek}
\usepackage{hhline}
\usepackage{soul}
\usepackage{url}

\def\um{\upmu\mbox{m}}

\begin{document}
\title{Laser-driven high-quality positron sources as possible injectors for plasma-based accelerators}

\author{A Alejo$^1$, R. Walczak$^2$, G Sarri$^1$}

\address{$^1$ Centre for Plasma Physics, School of Mathematics and Physics, Queen's University Belfast, University Road, Belfast BT7 1NN, UK\\$^2$ John Adams Institute for Accelerator Science and Department of Physics, University of Oxford, Denys Wilkinson Building, Keble Road, Oxford OX1 3RH, UK}
\ead{a.alejo@qub.ac.uk}

\begin{abstract}
The intrinsic constraints in the amplitude of the accelerating fields sustainable by radio-frequency accelerators demand for the pursuit of alternative and more compact acceleration schemes. Among these, plasma-based accelerators are arguably the most promising, thanks to the high-accelerating fields they can sustain, greatly exceeding the GeV/m. While plasma-based acceleration of electrons is now sufficiently mature for systematic studies in this direction, positron acceleration is still at its infancy, with limited projects currently undergoing to provide a viable test facility for further experiments. In this article, we propose a recently demonstrated laser-driven configuration as a relatively compact and inexpensive source of high-quality ultra-relativistic positrons for laser-driven and particle-driven plasma wakefield acceleration studies. Monte-Carlo simulations show that near-term high-intensity laser facilities can produce positron beams with high-current, femtosecond-scale duration, and sufficiently low normalised emittance at energies in the GeV range to be injected in further acceleration stages.   
\end{abstract}

\maketitle
\ioptwocol
\section{Introduction}

Since the experimental discovery of the positron in 1932~\cite{Anderson}, positron beams have played a major role in experimental physics, not only for fundamental studies but also for their wide range of practical applications, which include medicine and material science. Sub-relativistic or mildly relativistic positrons are naturally emitted by elements undergoing $\beta^+$-decay, a widely used source for positron annihilation spectroscopy and positron emission tomography. On the other hand, ultra-relativistic positron beams are widely used in experimental particle and nuclear physics, predominantly in electron-positron colliders. The current record for the highest centre-of-mass energy in an electron-positron collider belongs to the recently dismissed LEP at CERN, able to achieve up to 209 GeV. Recent proposals for the next generation of particle colliders aim at increasing this energy, eventually reaching the TeV-scale centre-of-mass energy. For example, the International Linear Collider (ILC)~\cite{ILC} and the Compact Linear Collider (CLIC)~\cite{CLIC} aim, in their first stage of implementation, to achieve 250 GeV and 380 GeV, respectively.

Increasing collider energies to hundreds of GeV would allow precision studies of the Higgs boson and the top quark, such as the possibility of the Higgs boson being a composite particle \cite{Higgs1} and its coupling with the top quark and itself \cite{Higgs2,Higgs3}, besides providing hints as to why it condenses giving a non-zero vacuum expectation value \cite{Higgs4}. Breaking the TeV barrier, on the other hand, would allow for searches beyond the Standard Model of particle physics. Besides the challenging requirement of accessing TeV-scale centre-of-mass energies, it is also necessary to create collisions at extremely high luminosity, since the typical cross sections involved scale as the inverse of the energy squared. For this reason, ILC and CLIC are aiming for luminosities greater than $10^{34}$ cm$^{-2}$s$^{-1}$, well beyond the $10^{31}$ cm$^{-2}$s$^{-1}$ previously achieved at LEP.

Conventional electron-positron colliders rely on established radio-frequency technology, which guarantees a maximum accelerating gradient at the level of tens to a $100\,$MV/m. This field is limited by the dielectric breakdown of the materials involved, implying that TeV energies can only be reached after tens to hundreds of km of acceleration. The sheer scale of these machines is thus a fundamental limiting factor in the development of lepton colliders justifying research into alternative acceleration schemes that can provide similar performance but in a more compact configuration. Among many, plasma-based wakefield acceleration is arguably one of the most promising, since it can sustain accelerating fields well beyond the GeV/m \cite{LWFAreview}, with experimentally demonstrated fields up to 100 GeV/m \cite{Liu}. Recent promising results in this area include, for instance, the demonstration of energy doubling of a 42 GeV electron beam in less than one meter of plasma \cite{Blumenfeld}, a 2 GeV energy gain of a positron beam in one metre of plasma \cite{Corde}, and the laser-driven acceleration of electrons up to 4.2 GeV in only 10 cm of plasma \cite{Leemans}. Thanks to its fast-paced development, particle-driven wakefield acceleration is now actively pursued in large-scale projects, including AWAKE at CERN \cite{AWAKE} and FACET at SLAC \cite{FACET} (together with its current upgrade, FACET-II \cite{FACETII}). On the other hand, laser-driven wakefield is currently studied only in relatively small University-scale laboratories even though the physics of laser-driven wakefield acceleration is in principle sufficiently mature to widespread its use to larger-scale applications. One example in this direction is the European funded project EuPRAXIA \cite{EuPRAXIA}, which aims at building a plasma-based electron accelerator of industrial quality able to accelerate, in a stable and consistent manner, narrow-band and high-current electron beams with a maximum energy of 5 GeV.

Developing plasma based accelerator technology to the level required by TeV-scale colliders represents a major challenge requiring progress in many fields as outlined in the recently published US  roadmap for advanced accelerators~\cite{USroadmap}. In a nutshell, a possible proposal is to use multi-staged wakefields, following promising results in proof-of-principle experiments of double-staged electron acceleration~\cite{Steinke}.

Even though the field of plasma-based electron acceleration is advancing fast, experimental studies of positron acceleration in a plasma is of a more challenging nature, due to the difficulty of providing an injector of suitable quality that can be, for instance, synchronised with the positron-accelerating region of a wakefield. 
To date, only the proposed upgrade of FACET \cite{FACETII} will be able, in the near future, to provide a source of ultra-relativistic positrons suitable for test studies of subsequent wakefield acceleration. In preliminary studies, including those in FACET-I~\cite{FACET}, two alternative methods have been studied. The first option uses an electron beam as a driver and a positron beam as a witness~\cite{Lotov}, both propagating inside a hollow channel to avoid the positron-defocussing fields~\cite{Schroeder}. The temporal synchronisation is ensured by propagating the electron drive through a thin converter to generate the positron beam~\cite{Wang}. Alternatively, the self-loaded plasma wakefield acceleration has been proposed~\cite{Corde, doche}, in which a high-charge positron beam is propagated through a hollow channel. The front of the positron bunch is in charge of generating the wakefield, whose accelerating fields are experienced by the back of the bunch. %COMMENT: need to comment on the PRL by Lindstrom

Providing positron beams of suitable quality to study these acceleration schemes is indeed extremely challenging, with effectively only FACET-II as a viable facility proposed to date. However, recent experimental results on laser-based generation of high-quality ultra-relativistic positrons are suggesting an alternative pathway towards this goal. For instance, G. Sarri and co-authors have recently reported on the generation of fs-scale and narrow divergence positron beams in a plasma-based configuration \cite{SarriPRL,SarriNCOMM}. In a nutshell, the positrons are generated as a result of a quantum cascade initiated by a laser-driven electron beam propagating through a high-Z solid target. For sufficiently high electron energy and thin converter targets, the generated positrons present properties that resemble those of the parent electron beam, hence the fs-scale duration~\cite{Lundh}, mrad-scale divergence~\cite{Osterhoff}, and small source size~\cite{Kneip, Brunetti}. The maximum positron energy attainable in this scheme is naturally dictated by the peak energy of the parent electron beam. Positrons with energy up to 0.5 GeV were produced in recent experiments~\cite{SarriNCOMM, SarriPPCF2}. These beams present unique advantages for being injected in further wakefields, when compared to more conventional sources. For instance, they have durations comparable to the positron-accelerating region of a wakefield~\cite{Wang} and are naturally synchronised with a high-power laser. However, other characteristics still need to be carefully optimised, such as their non-negligible normalised emittance and the relatively low charge. Providing a high-quality source of ultra-relativistic positrons is however of paramount importance for the development of plasma-based accelerators of positrons, to date an area of research predominantly of a theoretical nature (see, for instance, Ref. \cite{positrons_wakefield}). 

In this paper, we report on an extensive numerical study devoted to define precisely the main properties of laser-driven positrons and assess their suitability to be further transported and manipulated in order to act as a seed for further wakefield-based acceleration stages. In this study, we assume three main ranges of parameters for the primary electron beam. In the first range, we consider typical electron parameters obtainable by 100\,TW commercially-available laser systems (maximum electron energy of 1 GeV). In the second, we assume state-of-the-art electron beam properties, similar to the characteristics reported in Ref. \cite{Leemans} and planned to be achieved by EuPRAXIA \cite{EuPRAXIA} (maximum electron energy of 5 GeV). Finally, we consider near-term beam properties expected to be achieved by the next generation of ultra-high intensity laser systems, such as the Extreme Light Infrastructure Nuclear Pillar \cite{ELI-NP} (maximum electron energy of 20 GeV). Our simulations show that high current (exceeding the kA) positron beams with sufficiently low emittance and high density can be generated, providing a firm baseline for the design of a test facility dedicated to optimising positron beam manipulation and further acceleration.

\section{Spatial and spectral properties of the positrons at source}
\begin{figure}[b!]
	\centering
	\includegraphics[width=\linewidth]{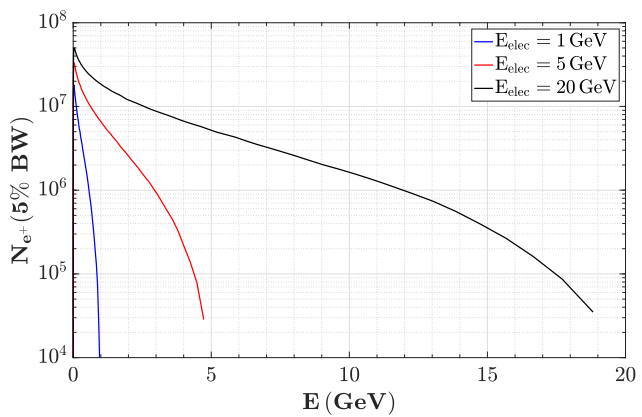}
	\caption{\textbf{Positron spectra}. Spectra of the positrons escaping a 1cm lead target traversed by an electron beam carrying 100 pC of charge and with an energy bandwidth of 5\%. Different primary electron energies are considered: 1 GeV (blue), 5 GeV (red), and 20 GeV (black). The spectra show the number of positrons as a function of energy in a 5\% bandwidth.}
	\label{spectra}
\end{figure}

In order to assess the main characteristics of the positron beam, we simulated the propagation of an ultra-relativistic electron beam through a high-Z solid target, using the Monte-Carlo scattering code FLUKA \cite{FLUKA1, FLUKA2}. For its practical simplicity of use, we simulate lead as a converter target and we assume a length of the order of two radiation lengths ($2L_{RAD}\approx 1$ cm for lead). This length is the one that maximises the number of escaping positrons, for any defined set of parameters of the incoming electron beam \cite{SarriPRL,SarriPPCF}. In order to make the obtained characteristics of the positron beam independent from those of the parent electrons, we assume a pencil-like electron beam with a point-like source and zero temporal duration. For a realistic electron beam, it is thus only necessary to convolute the parameters of the parent electrons with those of the positrons. For each simulation, $10^6$ primary electrons are assumed, and consider the positrons exiting from the rear side of the converter. The main properties of the positron beams that we consider are: source size $d$, divergence $\theta_{div}$, temporal duration $\tau$, Lorentz factor $\gamma$, geometrical emittance $\epsilon$, and normalised emittance $\bar{\epsilon}$. As mentioned in the introduction, three different electron energies are chosen: $E_0 =$ 1, 5, and 20 GeV. For each energy, we simulated a primary electron beam with a Gaussian spectrum centred around $E_0$ and a standard deviation $\sigma_E = 5\%E_0$. These three energies are chosen in order to represent the typical performance of a laser-driven electron accelerator employing a commercial laser system, state-of-the art high-intensity lasers \cite{Leemans}, and the next generation of high-intensity lasers (such as ELI \cite{ELI-NP}), respectively. For the 5 GeV case, it is also worth noticing that this is the baseline energy of the EuPRAXIA project \cite{EuPRAXIA}. For the number of positrons escaping the solid target, we generally assume that the primary electron beam carries a charge of 100 pC. However, the number of positrons is directly proportional to the number of electrons in the parent beam, allowing for a simple rescaling of the results for different initial electron charges. %Moreover, we emphasize that the choice of a narrow energy spread in the primary electron beam is not of fundamental importance for the characteristics of the generated positrons. Their spectrum is predominantly dictated by the cascade inside the solid, virtually independent from the original electron spectrum.

The spectra of the positrons escaping the converter are depicted in Fig. \ref{spectra}. As expected from the quantum cascade in the solid, the spectra present a monotonic shape with a maximum energy corresponding to the energy of the parent electron beam \cite{SarriPPCF}. However, it is worth noticing that a 100 pC electron beam, as routinely obtainable in a laser wakefield accelerator, is able to generate a significant number of positrons, even in a narrow energy slice. For instance, assuming EuPRAXIA-like parameters, up to $10^{11}$ positrons with energy exceeding 100 MeV can be generated, with up to $6\times10^6$ positrons in a 5\% bandwidth around 1 GeV, corresponding to a charge of approximately 1 pC. This number raises to $1.8\times10^7$ (charge of 3 pC) if an initial electron beam with an energy of 20 GeV is considered. As a further comment, it must be noted that the spectrum of the generated positrons is virtually insensitive to the spectrum of the parent electron beam. It is thus preferable to use high-charge broadband electron beams, rather than focussing on narrow-band laser-wakefield acceleration. If we assume a 1 nC electron beam with a broadband spectrum up to 5 GeV, we can then obtain up to 10 pC of positrons in a 5\% bandwidth around 1 GeV. 
\begin{figure}[b!]
	\centering
	\includegraphics[width=\linewidth]{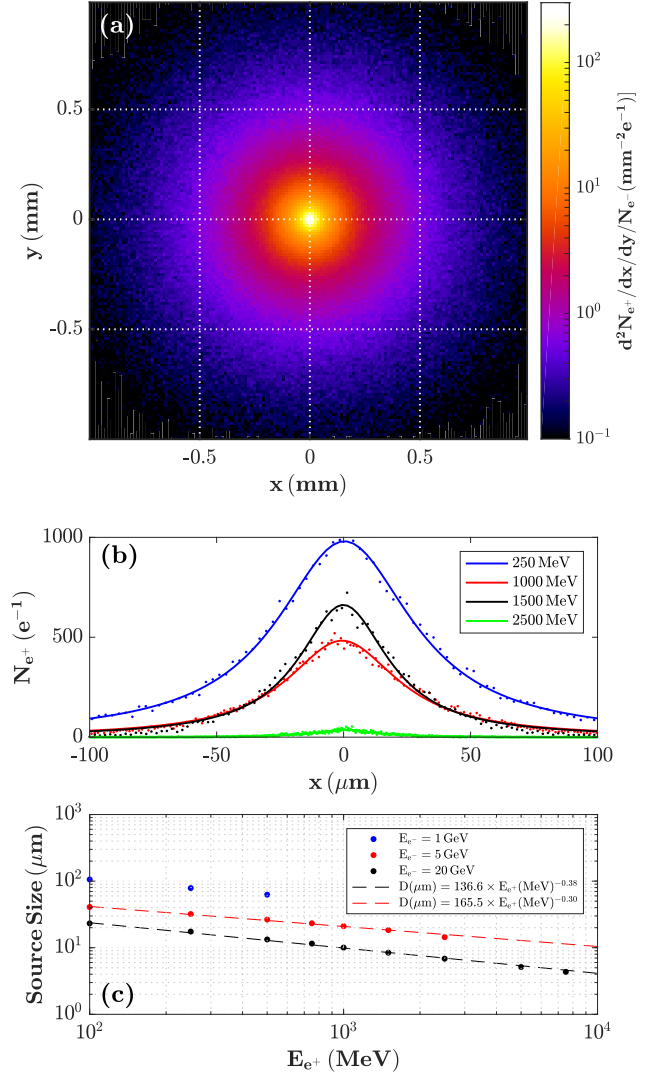}
	\caption{\textbf{Energy-dependent source-size of the positron at the exit of the converter target}. \textbf{(a)} Spatial distribution of the entire positron beam at the rear surface of the converter. \textbf{(b)} Transverse distribution of the positrons for different positron energies, generated by a $5\pm5\%\,$GeV electron beam, fitted by Lorentzian distributions of different width. \textbf{(c)} Extracted energy-dependent size of the real source of the positrons for different parent electron beam energies.  The source size of the positrons is seen to scale approximately as the cubic root of their energy (as fitted by the red and black dashed lines).}
	\label{source_size}
\end{figure}

This is a sizeable positron beam that could in principle be injected in further stages of laser-driven or particle-driven wakefield acceleration, provided that the beam possesses sufficient spatial quality. In particular, it is necessary to assess the source size, the divergence, and the emittance of the beam. Due to the generation of the positrons throughout the converter, the definition of the source is somewhat ambiguous. In our case, we define the real source size as the size of the positron beam at the rear surface of the converter and a virtual source size corresponding to the waist of the beam obtained by back-tracking the generated positrons. 

The positron beam profile at the exit of the converter target is depicted in Fig.~\ref{source_size}(a) for the case of a 5 GeV initial electron beam, accounting for all the positrons emitted independent of their energy or time of arrival. As it can be seen, the positron beam exhibits a smooth profile, in agreement with previous experimental observations~\cite{SarriPPCF2}. The source size of the overall beam can be estimated to be $\sim33\um$, given by the radius or Half-Width-at-Half-Maximum of the distribution. However, the source size is energy-dependent with the spatial distribution being well-approximated by a Lorentzian distribution: 
$N_{e^+}\simeq k/\left[\pi\sigma_x\left(1+\left(x/\sigma_x\right)^2\right)\right]$ (see Fig. \ref{source_size}(b)), with smaller sizes ($\sigma_x$) for higher energies (see Fig. \ref{source_size}(c)). As an example, 1 GeV positrons have a source size of around 20 microns (10 microns) if a parent 5 GeV (20 GeV) electron beam is assumed.  The virtual source size is found to be about 16 micron (7.5 micron), placed inside the converter at a distance of 1.6mm (0.8mm) from its rear surface. This source size arises from assuming a pencil-like electron beam with a point-like source. In order to compute the realistic positron source  size one would then need to convolute these results with the intrinsic beam size of the specific electron beam used. However, laser-driven electron beams typically have a source size in the micron range \cite{Kneip} and typical divergences in the mrad range \cite{Osterhoff}, inducing only small corrections to the results reported here.

\begin{figure}[t!]
	\centering
	\includegraphics[width=\linewidth]{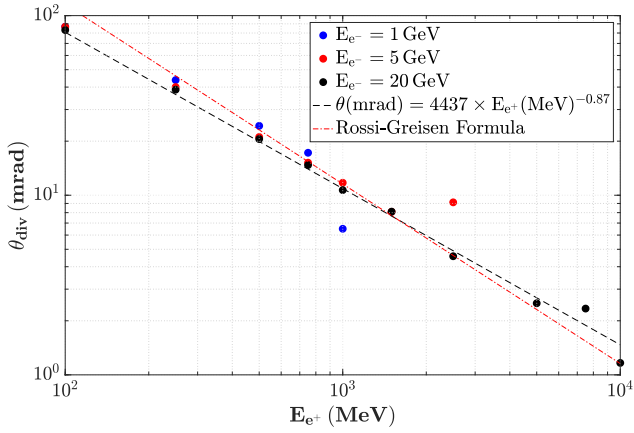}
	\caption{\textbf{Energy-dependent divergence of the positrons at the exit of the converter target.} Different colours correspond to different energies of the primary electron beam. At multi-GeV, the divergence of the positrons scales approximately as the inverse of their energy (fit shown by the black dashed line), in good qualitative agreement with the Rossi-Greisen formula (red dashed line).}
	\label{divergence}
\end{figure}

Similarly, the divergence of the positrons is seen to decrease for increasing positron energies, as depicted in Fig.~\ref{divergence}. At 1 GeV the positrons have a divergence of the order of 10 mrad, which decreases further down to 3 mrad at 5 GeV. The divergence is seen to scale with the positron energy as $\propto E^{-0.87}$, in good qualitative agreement with the Rossi-Greisen formula~\cite{Greisen} routinely used in accelerator physics. Again this divergence should be convoluted with the divergence of the primary electron beam, inducing a small correction, in the mrad range.

\section{Temporal properties of the positrons at source}

\begin{figure}[b!]
	\centering
	\includegraphics[width=\linewidth]{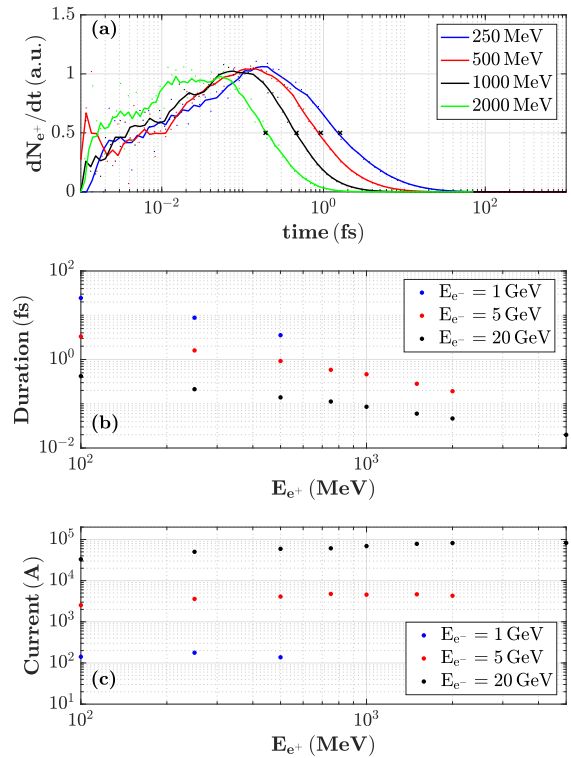}
	\caption{\textbf{Temporal properties of the positron beam at source.} \textbf{(a)} Temporal profile of the positrons of energy 250, 500, 1000 and 2500 MeV, produced from the interaction of a 5 GeV electron beam with a 1cm-thick Pb target. The points are well fitted by a log-normal distribution. \textbf{(b)} Duration of the positron beam as a function of their energy. \textbf{(c)} Peak current of the positrons contained within each 5\% energy bin.}
	\label{current}
\end{figure}

A defining feature of laser-driven electron beams is their short temporal duration, down to even the femtosecond-scale~\cite{Lundh}. Such a short duration is in principle achievable by radiofrequency accelerators, but at the cost of implementing complex beam slicing devices~\cite{beam_slicing}. It is thus to be expected that positrons generated from such electron beams will present similarly short durations. In order to extract the temporal duration of the positron beam at source, a specific add-on code was written in the source files of FLUKA. Obtained beam durations are shown in Fig. \ref{current}. Frame a) depicts the temporal distributions of positrons of different energy. Each distribution resembles a log-normal distribution with different variances. 
%Each distribution is reasonably well fitted by a log-normal distribution with different variances. 

As intuitively expected, higher positron energies correspond to shorter beam durations, of the order of approximately 1 fs (longitudinal size of 300 nm) at 1 GeV for a 5 GeV parent electron beam. This effectively implies that the 1 GeV positron beams will have virtually the same duration as the primary electron beam. Assuming a realistic electron beam with a duration of 5fs, we obtain a longitudinal size of the positron beam of the order of $\sigma_z\approx 1.5\,\upmu$m. 

This value is much smaller than the one proposed for FACET-II \cite{FACETII} (10 $\upmu$m) and indeed extremely useful for high-energy applications. For instance, a short longitudinal size is highly desirable for injection in additional wakefield acceleration stages. The extremely small positron-accelerating region, of a few microns, in plasma wakefield accelerators have led to alternative schemes being investigated, such as self-loaded PWFA~\cite{corde, doche}. In SL-PWFA, the front of a long, high-charge positron beam drives a wakefield, which accelerates the positrons contained in the back of the bunch, in a positron-accelerating region larger than in the regular electron-driven wakefield. However, such a mechanism requires very large positron charges in order to generate the wakefield, while only a small fraction of these positrons are actually accelerated. Furthermore, the extension of the accelerating region comes at the cost of a significant reduction of the accelerating field. The intrinsically-short durations of laser-driven positron beams would allow the positron acceleration in the original wakefield produced by a laser or an electron drive beam, reducing the number of stages required for high-energy applications.

Moreover, a short beam duration is recommended in order to minimise beam disruption in high-energy electron-positron collisions, as caused by beam focussing induced by the electromagnetic fields of the other beam. The beam disruption parameter is in fact directly proportional to the beam longitudinal size~\cite{chen1988}: 
\begin{equation}
D_{x,y} \approx \frac{2r_eN\sigma_z}{\gamma \sigma_{x,y},(\sigma_x+\sigma_y)},
\end{equation}
where $r_e$ is the classical electron radius, $\gamma$ is the particle Lorentz factor, $N$ is the number of particles in a beam, and $\sigma_{x,y}$ are the transverse dimensions of the beam.

Moreover, a short longitudinal size of the beam is also desirable in order to minimise beamsstrahlung, one of the main sources of background noise in high-energy colliders~\cite{schroeder2012}. In its quantum regime, the number of photons generated via beamsstrahlung scales as $n_{phot}\propto\sigma_z^{1/3}$~\cite{delahaye1999} and the number of pairs produced coherently $n_{coh}\propto\sigma_z^{2/3}$~\cite{yokoya1990}

Finally, a short beam duration naturally translates into a high peak current of the beam (Fig. \ref{current}.c). In a 5\% slice centered around 1 GeV, a 5 GeV, 100 pC electron beam can generate a positron current exceeding the kA. Even though one must note that a high current is not explicitly required in liner colliders, this characteristic is still of great interest for other applications in material science.

\section{Emittance of the positrons at source}

\begin{figure}[h!]
	\centering
	\includegraphics[width=\linewidth]{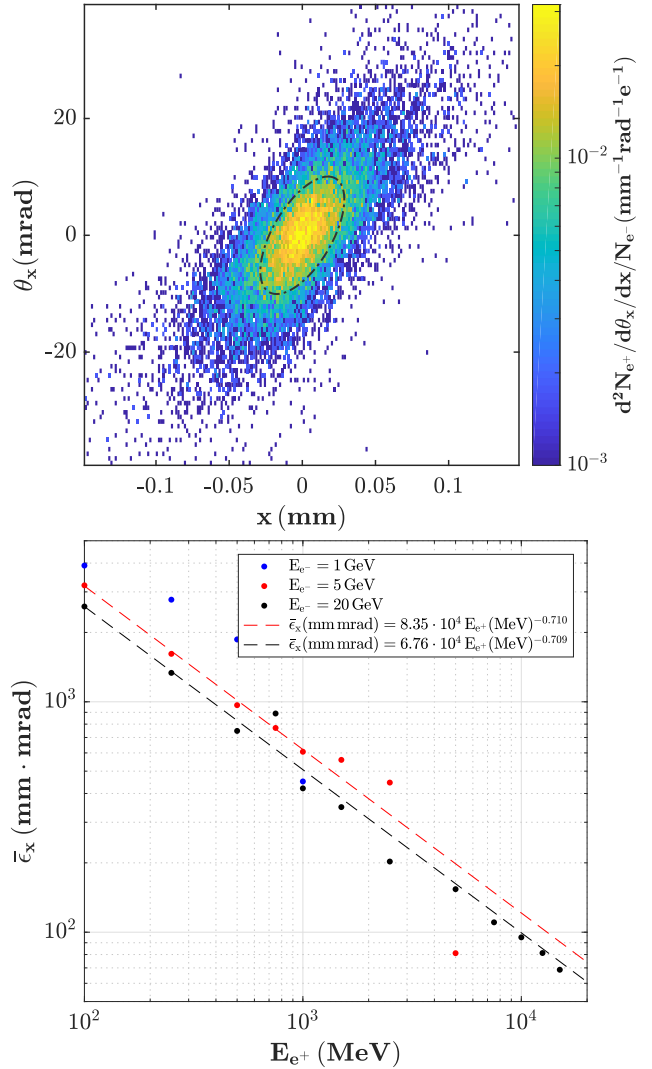}
	\caption{\textbf{Positron emittance}. \textbf{(a)} Phase-space diagram of 1 GeV positrons exiting the converter target. \textbf{(b)} Energy-dependent normalised emittance of the positron at the exit of a 1-cm thick Pb target for different initial electron energies. The emittance of the positrons scales approximately as $E^{-2/3}$.} 
	\label{norm_emittance}
\end{figure}

The positrons generated inside the converter are emitted in a beam-like fashion. Such a beam can be described, in a first approximation, by its source size and divergence. However, the most important parameter to characterise the spatial quality of a beam is arguably its emittance.
The emittance is a measure of the volume of the particle beam in phase-space, and defines the ease for beam transportation using magnetic fields, as well as being the main parameter in conjunction with the current defining the brightness of the beam. The emittance is mathematically defined as $\epsilon_x = \sqrt{\langle x^2\rangle \langle x'^2 \rangle - \langle xx' \rangle^2}$, where $x$ is the position of the particles and $x'$ its divergence. Typically, the emittance is defined for particles within a narrow energy range (monoenergetic approximation) and low angular divergence (paraxial approximation), which allows to define the normalised emittance as $\bar{\epsilon} = \gamma \beta \epsilon$, where $\gamma$ is the particle Lorentz factor and $\beta$ its velocity normalised to the speed of light in vacuum. 
As an example, the phase-space of positrons of energy $1.00\pm0.05\,$GeV leaving the converter is shown in Fig.~\ref{norm_emittance}(a), as generated from the interaction of a $5\,$GeV electron beam. The emittance in this case corresponds to the area of the ellipse defined by the rms of the distribution, depicted in Fig.~\ref{norm_emittance}(a) by the black, dash-dotted ellipse. With that, a geometrical emittance of $\epsilon_x=0.1\pi\,$mm mrad is obtained (normalised emittance $\bar{\epsilon}_x=200\pi\,$mm mrad). 
Given the strong dependence of both source size and divergence on the positron energy considered, the emittance is expected to follow a similar trend. The scaling of the normalised emittance with the positron energy bin is shown in Fig.~\ref{norm_emittance}(b), showing the data points to close follow a power-law function.

\section{Discussion}

The main properties of positrons generated from laser-driven electrons have been shown in the previous sections. These characteristics can be directly compared with those obtainable with more conventional generation schemes and with the requirements for injection in laser-driven and particle-driven wakefield acceleration stages. We take FACET-I and FACET-II as representative examples of state-of-the-art particle-driven wakefield facilities and this comparison is shown in table~\ref{tab:Comparison}.

%\begin{table}
%	
%	%\begin{center}
%	\footnotesize
%			\begin{tabular}{c|c|c|c|c|c}
%		 & Units & ILC & CLIC & FACET-II & LWFA \\
%		\hline\hline
%		$E$ & GeV & 500 & 380 & 10 & 1  \\
%	        $P$ & W & $10^7$ & $28\times10^6$ & 50 & 3  \\
%		$N_e$ & / & $2\times10^{10}$ & $3\times10^9$ & $6\times10^9$ & $6\times10^6$  \\
%		$\sigma_x$ & $\mu$m & 0.5 & 0.04 & 10 & 10\\
%		$\sigma_y$ & $\mu$m & $6\times10^{-3}$ & $10^{-3}$ & 10 & 10\\
%		$\sigma_z$ & $\mu$m & 300 & 44 & 10 & 0.6\\
%		$\epsilon^*_x$ & $\mu$m & 10 & 0.66 & 1 & 500\\
%		$\epsilon^*_y$ & $\mu$m & 0.035 & 0.02 & 1 & 500\\
%		$\Delta E$ & \% & 0.1 & 0.3 & 1 & 1 \\
%		$N_{B}$ & / & 1300 & 300 & 1 & 1 \\
%		$ f $ &  Hz & 5 & 50 & 1 & $10^{3}$ \\
%		$\ell$ & cm$^{-2}$s$^{-1}$ & $1.8\times10^{34}$ & $6\times10^{34}$ & $3\times10^{25}$ & $10^{24}$ \\
%		
%		\end{tabular}
%	%\end{center}
%	\label{table1}
%	\caption{bla bla bla} 
%\end{table}

\begin{table}[b!]
\centering
\begin{tabular}[b!]{ccccc}
\hhline{=====}
		 & Units & FACET-I & FACET-II & LWFA  \\
\hline\hline
		$E$ & GeV  & 21 & 10 & 1  \\
	    $P$ & W & 7.4 & 9.6 & 3  \\
		$Q_e$ & pC & 350 & 500 & 2  \\
		$\sigma_x$ & $\upmu$m& 30 & 4 & 16\\
		$\sigma_y$ & $\upmu$m& 30 & 4 & 16\\
		$\sigma_z$ & $\upmu$m& 50 & 6.4 & 0.6\\
		$\bar{\epsilon}_x$ & mm\,mrad & 200 & 7 & 500\\
		$\bar{\epsilon}_y$ & mm\,mrad & 50 & 3 & 500\\
		$\Delta E$ & \% & 1.5 & 1 & 5 \\
		%$N_{B}$ & / & 1 & 1 & 1 \\
		$ f $ &  Hz & 1 & 1 & $10$ - $10^{3}$ \\
		$\ell$ & cm$^{-2}$s$^{-1}$ & $5\times10^{23}$ & $6\times10^{25}$ & $10^{22 - 24}$ \\
\hhline{=====}
\end{tabular}
\caption{\textbf{Comparison with FACET-I and FACET-II} the main positron beam parameters obtained in FACET-I and expected for FACET-II \cite{Joshi} are compared with typical parameters at source from a laser wakefield scheme as discussed in this manuscript. The latter assumes a 100 pC 5 GeV primary electron beam. $E$ denotes the beam energy, $W$ the average power, $Q_e$ the total charge (in a 5\% bandwidth centered around 1 GeV in the LWFA case), $\sigma_{x,y}$ the transverse sizes, $\sigma_z$ the longitudinal size, $\bar{\epsilon}_{x,y}$ the normalised emittance, $\Delta E$ the energy spread, $f$ the frequency of operation, and $\ell$ the luminosity.}
\label{tab:Comparison}
\end{table} %COMMENT: let's discuss the details of this table

As one can see, the main drawback for the use of laser-driven sources is given by the lower number of positrons in the beam, limited to a few pC of charge for realistic charges of the parent electron beam (100pC-1nC). This charge is two orders of magnitude lower than that attainable with FACET, and up to three orders of magnitude lower than the charges in conventional accelerators, such as the Stanford Linear Collider (SLC, 1.6nC), Large Electron Positron collider (LEP, 240pC), or estimated for the International Linear Collider (ILC, 3.2nC). The main reason for the lower charge is the significantly lower electron charge attainable via LWFA compared to the 10s-100s nC of charge in conventional accelerators in the other cases. However, the positron beams from conventional accelerators typically form bunches significantly longer than those for the case of laser-driven sources. Thus, the positron densities for a ns-long from a conventional accelerator would be of the order of a few Coulomb per second, lower than the hundreds of Coulombs per second in the case of beams with a few pC in a few fs for the laser-driven sources. Furthermore, it should be noted that few pC charges is sufficient for most applications (including cooling and acceleration studies for positrons towards colliders), particularly with the advances in high power lasers towards higher repetition rates.

As one can see, the main drawback for the use of laser-driven sources is given by the lower number of positrons in the beam, limited to a few pC of charge for realistic charges of the parent electron beam (100pC-1nC). This charge is two orders of magnitude lower than that attainable with FACET, and up to three orders of magnitude lower than the charges in conventional accelerators, such as the Stanford Linear Collider (SLC, 1.6nC), Large Electron Positron collider (LEP, 240pC), or estimated for the International Linear Collider (ILC, 3.2nC). The main reason for the lower charge is the significantly lower electron charge attainable via LWFA compared to the 10s-100s nC of charge in conventional accelerators in the other cases. However, it should be noted that few pC charges is sufficient for most applications (including cooling and acceleration studies for positrons towards colliders), particularly with the advances in high power lasers towards higher repetition rates.
The main bottleneck to reach higher repetition rates in high power lasers has been the cooling of the lasing medium, due to the use of white flashlamps as pumps. The use of diodes with wavelength in the absorption band for the laser emission will allow for higher repetition rates. For instance, the DiPOLE project has recently demonstrated a stable system delivering 10 J laser pulses at 10 Hz. The design is scalable and demonstration of 100 J laser pulse energy at 10 Hz is forthcoming, eventually aiming for a 1kJ 10 Hz laser. The need for the development of 1kHz high power lasers is also identified in the US roadmap for future novel accelerators.
Such high repetition rates will help compensating the lower single-shot charge by providing a greater time-averaged charge. 

Furthermore, the total charge in the beam can be increased significantly simply by considering a larger energy bandwidth. For example, the final stage of LEP used positrons accelerated by the pre-linac with energy in the range $90\pm 7\%$\,MeV, and SLC could accelerate up to $20\%$ of the positrons initially contained in the energy range $11\pm 80\%$\,MeV. Considering both solutions, the time-averaged charge of LWFA-driven positrons would reach values of hundreds of pC per second, comparable to that of the FACET projects.

As shown in table~\ref{tab:Comparison}, the normalised emittance of the LWFA-based positrons is already comparable to that of FACET-I. Such an emittance would thus allow for a direct injection in a particle-based wakefield accelerator, without the need of a linear accelerator and emittance damp to produce the positron beam. In fact, recent studies~\cite{doche} have shown that a positron beam with degraded emittance can still be significantly accelerated using PWFA in FACET-I, at the price of achieving a lower energy gain. With respect to conventional accelerators, the emittance of the LWFA-based positrons is significantly lower than that in their initial acceleration stages. For example, the initial stages of SLC presented normalised emittances of $6500\pi$ mm\,mrad (geometrical emittance of 500 mm\,mrad), and  $11200\pi$ mm\,mrad (geometrical emittance of $62\pi$ mm\,mrad) for the case of LEP. The largely reduced emittance would allow for the injection in conventional accelerators without the need of emittance damps, that would reduce the charge in the beam and increase the cost and size of the system. 

Finally, the temporal duration of the positrons by laser-driven sources is significantly shorter than the other sources. Conventional accelerators naturally produce beams with durations of the order of a few ns, which can then be treated to reduce it down to hundreds of fs, at the cost of a reduction in the charge in the bunch. The shorter duration, automatically given by the intrinsically short duration of the electron beam, presents direct advantages for applications in material science. Additionally, as mentioned earlier, the reduced longitudinal dimension of the beam is smaller than the positron-accelerating region of the wakefield in a plasma~\cite{Wang}. The region of positron accelerating and focussing fields in an electron- or laser-driven plasma wakefield is limited to a sub-10 micron region along the propagation axis ($\sim35\,$fs).
Although future studies, both in particle-driven or laser-driven wakefields in FACET-II and EuPRAXIA, might open up new possibilities for a more robust and flexible positron acceleration scheme, currently only short beams fitting that region, such as the laser-driven positrons, can be efficiently accelerated by the wakefield. Furthermore, the laser-driven positrons present additional advantages. For instance, the wakefield structure can be significantly modified if a high-charge positron beam is used as a witness~\cite{Lotov}, reducing the peak accelerating field. Although high charges are desirable for schemes such as self-loaded PWFA, in which the positron beam both generates the wakefield and is accelerated~\cite{Corde}, lower charges may be preferred in more conventional plasma wakefield accelerators to maximise the electric field and reduce the number of stages required. Finally, a laser-driven source would also present the advantage of being jitter-free, allowing a stable synchronisation on a fs-scale. This problem is partially solved by using the drive electron beam to generate the positrons by propagating through a high-Z material~\cite{Wang}. However, such a procedure results in a slight worsening of the electron drive beam emittance, as well as adding the technological complication of ensuring the appropriate temporal separation between the electron drive and the positron witness beams, to ensure the synchronisation between the positron beam and the wakefield rather than the electron beam. A laser-driven source, on the other hand, would be intrinsically synchronised with a high-power laser. The lack of jittering, thanks to the laser beams being produced from the same oscillators, ensures a stable synchronisation of the different laser-driven wakefield stages, with temporal matching to the fs-level already achieved experimental~\cite{Corvan}. It should be noted, however, that in the case of TeV accelerators, beam cooling might be required in order to focus the beams to the nanometer-scale, boosting the overall luminosity. Further studies would be required regarding the cooling needs, which would come at the cost of extending the bunch longitudinal size (duration), but would allow for a broader bandwidth accepted, significantly increasing the number of positrons in the beam.

A final note must be made regarding the transport and energy selection of the laser-driven positrons. As shown earlier, the $1.00\pm0.05\,$GeV positrons generated exhibit a low emittance. However, it should be noted that this energy band has to be selected from the broadband positron emission from the converter. 
Also, even the positrons contained in the reduced bandwidth present a significant level of divergence, rapidly expanding the beam to sizes greater than the extension of the positron-focussing fields in the wakefield, of the order of tens of microns. Therefore, the positron beam requires collimation and re-focussing in order to be further accelerated. Both collimation and energy-selection appear feasible using magnetic systems. Laser-driven schemes are compact and produce lower levels of radiation, allowing for closely-coupled experiments. Thanks to that, a compact quadrupole triplet or a plasma lens can be used to collimate the positron emission. Given the low divergence and narrow energy band considered, a large portion of the positron beam can be collected by the system, greater than the $20\%$ collected at SLC, in which a broader energy range was accepted. Once the beam has been collimated, well-known energy selection system based on magnetic chicanes, such as those in conventional accelerators, can be implemented to select the positron energies to be further transported.

\section{Conclusions}
We have simulated the main properties of laser-driven positrons in order to test the feasibility of their injection in a multi-stage, plasma-based accelerator. From the interaction of the LWFA-driven electrons with the high-Z converter, a broadband positron beam can be generated, whose characteristics significantly improve with energy. For an electron beam similar to that proposed by the EuPRAXIA project (5 GeV, 100pC), up to $5\times10^6$ positrons are generated in a energy range $1.00\pm0.05\,$GeV, with a normalised emittance of $190\pi$\,mm\,mrad. Such an emittance is compatible with the further injection on secondary acceleration stages of plasma-based accelerators. Additionally, the naturally short duration of the positron beam, similar to that of the parent electron beam, can present additional advantages towards the further acceleration in a wakefield structure, as well as being beneficial for future colliders thanks to the reduced beamsstrahlung. Laser-driven positron beams can thus constitute an appealing alternative  as a witness beam for beam-driven and laser-driven wakefield test facilities and, potentially, as injector of reduced cost and size for future high energy accelerators and lepton colliders.

\section{Acknowledgments}
The authors acknowledge financial support from EPSRC (Grant No: EP/N027175/1 and EP/P010059/1). This work was supported by the European Union's Horizon 2020 research and innovation programme under grant agreement No. 653782. The simulation data are available at \emph{(URL to be inserted)}


\section{References}

\begin{thebibliography}{100}
\bibitem{Anderson} C. D. Anderson, Phys. Rev. \textbf{43}, 491 (1933).
\bibitem{ILC} Behnke, Ties, et al. "The International Linear Collider Technical Design Report-Volume 1: Executive Summary." arXiv preprint arXiv:1306.6327 (2013).
\bibitem{CLIC} Aicheler, M., et al. A Multi-TeV linear collider based on CLIC technology: CLIC Conceptual Design Report. No. SLAC-R-985. SLAC National Accelerator Lab., Menlo Park, CA (United States), 2014.
\bibitem{Higgs1} M. J. Dugan, H. Georgi and D. B.Kaplan, Anatomy of a Composite Higgs Model, Nucl. Phys. B254, 299 (1985).
\bibitem{Higgs2} Alekhin, S., Djouadi, A., and Moch, S. (2012). The top quark and Higgs boson masses and the stability of the electroweak vacuum. Physics Letters B, 716(1), 214-219.
\bibitem{Higgs3} Baur, Ulrich, Tilman Plehn, and David Rainwater. "Measuring the Higgs boson self-coupling at the Large Hadron Collider." Physical review letters 89.15 (2002): 151801.
\bibitem{Higgs4} Higgs, Peter W. "Spontaneous symmetry breakdown without massless bosons." Physical Review 145.4 (1966): 1156.
\bibitem{LWFAreview} Esarey, E., C. B. Schroeder, and W. P. Leemans. "Physics of laser-driven plasma-based electron accelerators." Reviews of Modern Physics 81.3 (2009): 1229.
%\bibitem{100GVm} Tajima, T., and J. M. Dawson. "Laser electron accelerator." Physical Review Letters 43.4 (1979): 267.
\bibitem{Liu} Liu, J. S., et al. "All-optical cascaded laser wakefield accelerator using ionization-induced injection." Physical review letters 107.3 (2011): 035001.
\bibitem{Blumenfeld} I. Blumenfeld et al., Nature \textbf{445}, 741 (2007)
\bibitem{Corde} Corde, S\'ebastien, et al. "Multi-gigaelectronvolt acceleration of positrons in a self-loaded plasma wakefield." Nature 524.7566 (2015): 442.
\bibitem{Leemans} Leemans, Wim P., et al. "GeV electron beams from a centimetre-scale accelerator." Nature physics 2.10 (2006): 696.
\bibitem{AWAKE} \url{http://awake.web.cern.ch/awake/}
\bibitem{FACET} Hogan, M. J., et al. "Plasma wakefield acceleration experiments at FACET." New Journal of Physics 12.5 (2010): 055030.
\bibitem{FACETII} Yakimenko, Vitaly, et al. "FACET-II Accelerator Research with Beams of Extreme Intensities." (2016): TUOBB02.
\bibitem{EuPRAXIA} \url{http://www.eupraxia-project.eu/}
\bibitem{USroadmap} Advanced Accelerator Development Strategy Report: DOE Advanced Accelerator Concepts Research Roadmap Workshop. United States: N. p., 2016. Web. doi:10.2172/1358081.
\bibitem{Steinke} Steinke, S., et al. "Multistage coupling of independent laser-plasma accelerators." Nature \textbf{530},190 (2016)
\bibitem{Lotov} Lotov, K. V. "Acceleration of positrons by electron beam-driven wakefields in a plasma." Physics of plasmas 14.2 (2007): 023101.
\bibitem{Schroeder} Schroeder, C. B., D. H. Whittum, and J. S. Wurtele. "Multimode analysis of the hollow plasma channel wakefield accelerator." Physical review letters 82.6 (1999): 1177.
\bibitem{Wang} Wang, Xiaodong, et al. "Positron injection and acceleration on the wake driven by an electron beam in a foil-and-gas plasma." Physical review letters 101.12 (2008): 124801.
\bibitem{doche} Doche, A., et al. "Acceleration of a trailing positron bunch in a plasma wakefield accelerator." Scientific Reports 7.1 (2017): 14180.
\bibitem{SarriPRL} Sarri, Gianluca, et al. "Table-top laser-based source of femtosecond, collimated, ultrarelativistic positron beams." Physical review letters 110.25 (2013): 255002.
\bibitem{SarriNCOMM} Sarri, Gianluca, et al. "Generation of neutral and high-density electron–positron pair plasmas in the laboratory." Nature communications 6 (2015): 6747.
\bibitem{Lundh} Lundh, O., et al. "Few femtosecond, few kiloampere electron bunch produced by a laser–plasma accelerator." Nature Physics 7.3 (2011): 219.
\bibitem{Osterhoff} Osterhoff, Jens, et al. "Generation of stable, low-divergence electron beams by laser-wakefield acceleration in a steady-state-flow gas cell." Physical Review Letters 101.8 (2008): 085002.
\bibitem{Kneip} Kneip, S., et al. "Characterization of transverse beam emittance of electrons from a laser-plasma wakefield accelerator in the bubble regime using betatron x-ray radiation." Physical Review Special Topics-Accelerators and Beams 15.2 (2012): 021302.
\bibitem{Brunetti} Brunetti, E., et al. "Low emittance, high brilliance relativistic electron beams from a laser-plasma accelerator." Physical review letters 105.21 (2010): 215007.
\bibitem{SarriPPCF2} Sarri, G., et al. "Spectral and spatial characterisation of laser-driven positron beams." Plasma Physics and Controlled Fusion 59.1 (2016): 014015.
\bibitem{positrons_wakefield} Vieira, J., and J. T. Mendonça. "Nonlinear laser driven donut wakefields for positron and electron acceleration." Physical Review Letters 112.21 (2014): 215001.
\bibitem{ELI-NP} \url{http://www.eli-np.ro/}
\bibitem{FLUKA1} "The FLUKA Code: Developments and Challenges for High Energy and Medical Applications", T.T. Böhlen, F. Cerutti, M.P.W. Chin, A. Fassò, A. Ferrari, P.G. Ortega, A. Mairani, P.R. Sala, G. Smirnov and V. Vlachoudis, Nuclear Data Sheets 120, 211-214 (2014) 
\bibitem{FLUKA2} "FLUKA: a multi-particle transport code", A. Ferrari, P.R. Sala, A. Fasso, and J. Ranft, CERN-2005-10 (2005), INFN/TC\_05/11, SLAC-R-773
\bibitem{SarriPPCF} Sarri, Gianluca, et al. "Laser-driven generation of collimated ultra-relativistic positron beams." Plasma Physics and Controlled Fusion 55.12 (2013): 124017.
%\bibitem{Baier} V. N. Baier and V. M. Katkov, Pis?ma Zh. Eksp. Teor. Fiz. 88:2, 88 (2008).
%\bibitem{electronduration}
\bibitem{Greisen} Greisen, K. "The extensive air showers." Progress in cosmic ray physics 3.1 (1956).
\bibitem{beam_slicing} Emma, P., et al. "Femtosecond and subfemtosecond X-ray pulses from a self-amplified spontaneous-emission–based free-electron laser." Physical review letters 92.7 (2004): 074801.
\bibitem{corde} Corde, S\'ebastien, et al. "Multi-gigaelectronvolt acceleration of positrons in a self-loaded plasma wakefield." Nature 524.7566 (2015): 442.
\bibitem{chen1988} Chen, Pisin, and Kaoru Yokoya. "Disruption effects from the interaction of round e+ e− beams." Physical Review D 38.3 (1988): 987.
\bibitem{schroeder2012} Schroeder, C. B., E. Esarey, and W. P. Leemans. "Beamstrahlung considerations in laser-plasma-accelerator-based linear colliders." Physical Review Special Topics-Accelerators and Beams 15.5 (2012): 051301.
\bibitem{delahaye1999} Delahaye, Jean-Pierre, et al. "Scaling laws for e+/e− linear colliders." Nuclear Instruments and Methods in Physics Research Section A: Accelerators, Spectrometers, Detectors and Associated Equipment 421.3 (1999): 369-405.
\bibitem{yokoya1990} Yokoya, K. "Frontiers of Particle Beams: Intensity Limitations." Lecture Notes in Physics 400 (1990): 415.
\bibitem{Joshi} Joshi, C., et al. "Plasma wakefield acceleration experiments at FACET II." Plasma Physics and Controlled Fusion 60.3 (2018): 034001.
\bibitem{Corvan} Corvan, D. J., et al. "Optical measurement of the temporal delay between two ultra-short and focussed laser pluses." Optics express 24.3 (2016): 3127-3136.
\end{thebibliography}
\end{document}